\documentclass[12pt]{article}
\usepackage{graphicx}
\usepackage{amssymb}

\begin{document}

%
\def\papertitlepage{\baselineskip 3.5ex \thispagestyle{empty}}
\def\preprinumber#1#2#3{\hfill \begin{minipage}{4.2cm}  #1
                  \par\noindent #2
                  \par\noindent #3
                 \end{minipage}}
\renewcommand{\thefootnote}{\fnsymbol{footnote}}
%
%
\papertitlepage
\setcounter{page}{0}
\preprinumber{KEK Preprint 2002-135}{UTHEP-467}{hep-th/0302020}
\baselineskip 0.8cm
\vspace*{2.0cm}
\begin{center}
{\large\bf A Note on T-duality of Strings in Plane-Wave Backgrounds}
\end{center}
\vskip 4ex
\baselineskip 1.0cm
\begin{center}
            { Shun'ya~ Mizoguchi\footnote[2]{\tt mizoguch@post.kek.jp} }
\\
     \vskip -1ex
        {\it High Energy Accelerator Research Organization (KEK)} \\
     \vskip -2ex
        {\it Tsukuba, Ibaraki 305-0801, Japan} \\
     \vskip 2ex
     {Takeshi~ Mogami\footnote[3]{\tt mogami@het.ph.tsukuba.ac.jp}}
            \ \ {and} \
            { Yuji  ~Satoh\footnote[4]{\tt ysatoh@het.ph.tsukuba.ac.jp}}
\\
     \vskip -1ex
        {\it Institute of Physics, University of Tsukuba} \\
     \vskip -2ex
       {\it Tsukuba, Ibaraki 305-8571, Japan}

\end{center}
\vskip 10ex
%
\baselineskip=3.5ex
\begin{center} {\large\bf Abstract} \end{center}
We show, by direct computations of bosonic string spectra,
the $O(d,d; \mbox{\boldmath $Z$})$ $(d=1,2)$ T-duality
in the maximally supersymmetric IIB plane-wave background
compactified on $S^1$ and $T^2$. Only half of the ordinary set of zero
modes appear in the Hamiltonian. This ``half'' Narain lattice is proved
to
be stabilized by the T-duality group.
\vskip 2ex
%
%
%
%
%
\vspace*{\fill}
\noindent
February 2003
\newpage
\renewcommand{\thefootnote}{\arabic{footnote}}
\setcounter{footnote}{0}
\setcounter{section}{0}
\baselineskip = 0.6cm
\pagestyle{plain}

\section{Introduction}
Among string dualities, T-duality of flat toroidal compactifications is
best understood.  (See \cite{GPR} for a review.)  It typically contains
an
operation exchanging  a compactification radius $R$ with $\alpha'/R$
\cite{KY-SS}, which is a special element of the full
T-duality group $O(d,d;\mbox{\boldmath $Z$})$
\cite{N}  in the case of a $d$-torus.\footnote{We ignore the  trivial
\mbox{\boldmath $Z_2$} factor of $O(d,d;\mbox{\boldmath $Z$})$.}

T-duality is associated with target-space isometries.
The type IIB  maximally
supersymmetric plane-wave background \cite{BFHP1,BFHP2} has mutually
commuting spacelike Killing vectors  \cite{BFHP1}, and one naturally
expects
T-duality to hold also in this background. However, although one could
still
argue the classical equivalence, one needs some gauge fixing for
quantization,
and strictly speaking, in the presence of nontrivial Ramond-Ramond flux
the
direct comparison of the spectra has not been done in any covariant
formulation.
In light-cone gauge, the Green-Schwarz strings are known to be solvable
\cite{Met}, but then the world sheet theory becomes massive and that
renders the
known CFT proofs of T-duality unavailable. In this note, we show, by
direct
computations of string spectra in light-cone gauge,
the $O(d,d;\mbox{\boldmath $Z$})$ $(d=1,2)$ T-duality in the
IIB maximally supersymmetric plane-wave background compactified
on $S^1$ and $T^2$.

For the $S^1$ case, the mode expansions were already given in \cite{Mic}
and in its ``IIA dual" in \cite{AGGP}, though not in a form that can
easily be
compared with the IIB spectrum.\footnote{For a discussion of the strings
on $S^1$ on the IIB side, see also \cite{BBHIO}.}
  We will display them in such a way that
we may
examine duality of the two theories. A curious feature of the zero
modes will be
found; on the IIB side there are no momentum modes but appear only
winding
modes,
while on the IIA side both are present.  We will show that the winding
modes do not
contribute to the IIA Hamiltonian, and they are really dual. Also, for
the $T^2$ case,
we will give a proof that the Hamiltonian is invariant under the full
$O(2,2;\mbox{\boldmath $Z$})$ action.

Since we are particularly interested in the zero modes, we restrict
ourselves in this paper to the check of the bosonic spectra.
For the fermionic sector, the T-duality transformation rule of
Green-Schwarz strings was derived for $S^1$ compactification
to quadratic order in $\theta$ \cite{CLPS}
by using a generalized Buscher's method \cite{B} .
In plane-wave backgrounds in general, the Green-Schwarz action has been
shown \cite{RT-MMS} to truncate at quadratic order and take
the form given in \cite{CLPS,MT}. The invariance of the fermionic
spectrum may follow from the bosonic results and supersymmetry.

The notation that we use in this paper is as follows.
The maximally supersymmetric plane-wave background for type IIB strings
is
\begin{eqnarray}
&& ds^2 = 2dx^+dx^- - \mu^2 x^i x^i  (dx^+)^2 +dx^i dx^i,
     \nonumber \\
&& F_{5\;+ijkl} = 4\mu \epsilon_{ijkl},~~~
F_{5} = 0~~~\mbox{otherwise}. \label{noncompactpp}
\end{eqnarray}
$i,j,\ldots=1,\ldots,8$ are the indices for the transverse coordinates.
$\epsilon_{ijkl}$ is defined to be nonzero if
$\{i,j,k,l\}$ are the set \{1,2,3,4\} or \{5,6,7,8\}, and
$\epsilon_{1234}=\epsilon_{5678}=+1$.

It has 30 isometries \cite{BFHP1}, and among them
$k_{S^\pm_{ij}}$ $(i\neq j)$ given by
\begin{eqnarray}
k_{S^\pm_{ij}}&=&k_{e_i}\pm k_{e^*_j}, \nonumber\\
k_{e_i}&=&-\cos\mu x^+\partial_i -\mu x^i \sin\mu x^+
\partial_-,\nonumber \\
k_{e^*_i}&=&-\sin\mu x^+\partial_i +\mu x^i \cos\mu x^+  \partial_-
\end{eqnarray}
have a constant (unit) norm \cite{Mic}.

\section{$S^1$ compactification revisited}
We will first focus on the T-duality transformation along the isometry
$k_{S^+_{87}}$.
The coordinate system in which  this isometry is manifest is
\cite{Mic}
\begin{eqnarray}
x^+&=& X^+,\nonumber\\
x^-&=& X^- -\mu X^7X^8,\nonumber\\
x^I&=& X^I~~~(I=1, \ldots,6),\nonumber\\
\left[\begin{array}{c}x^7\\x^8\end{array}\right]&=&
\left[\begin{array}{rc}\cos \mu X^+&\sin\mu X^+ \\
-\sin\mu X^+&\cos\mu X^+\end{array}\right]
\left[\begin{array}{c}X^7\\X^8\end{array}\right].
\label{coordinates}
\end{eqnarray}
In these coordinates $k_{S^+_{87}}=-\partial/\partial X^8$, and the
metric becomes
\begin{eqnarray}
ds^2 &=& 2dX^+dX^- - \mu^2 X^I X^I  (dX^+)^2 -4\mu X^7 dX^8 dX^+ + dX^i
dX^i.
\end{eqnarray}
$F_{5}$ is unchanged.

\noindent{\it  IIB spectrum}\\
We will first examine the bosonic spectrum of a Green-Schwarz string in
this background \cite{Met,Mic,BBHIO}.
In light-cone gauge, the string action takes the form
\begin{eqnarray}
S_{IIB}&=&\int d\tau L_{IIB}, \nonumber\\
L_{IIB}&=&
\int_0^{2\pi} d\sigma
\left[
p^+\partial_\tau X^-
+\frac1{4\pi\alpha'}
\left(
(\partial_\tau X^i)^2-(\partial_\sigma X^i)^2
     -\tilde\mu^2 (X^I)^2 -4\tilde\mu X^7\partial_\tau X^8
\right)
\right],
\label{IIBstring}
\end{eqnarray}
where we have set
\footnote{The other gauge fixing conditions we have adopted are
$\det g_{\mu\nu}=-1$, $\partial_\sigma g_{\sigma\sigma}=0$
and $g_{\tau\sigma}(\sigma=0)=0$. Then $g_{\sigma\sigma}$ is constant
everywhere, and is set to be unity. }
$X^+=2\pi\alpha'p^+g_{\sigma\sigma}^{-1}\tau$ and
$\tilde\mu=2\pi\alpha' p^+\mu$.
The Hamiltonian is
\begin{eqnarray}
H_{IIB}&=&\frac1{4\pi\alpha'}\int_0^{2\pi}d\sigma\left[
(2\pi\alpha')^2
\left(
\Pi_I^2 + \Pi_7^2 +(\Pi_8 +\frac{\tilde\mu}{\pi\alpha'} X^7)^2
\right)
+(\partial_\sigma X^i)^2
+\tilde\mu^2 (X^I)^2
\right], \
\end{eqnarray}
where
\begin{eqnarray}
\Pi_I&=&\frac1{2\pi\alpha'}\partial_\tau X^I,\nonumber\\
\Pi_7&=&\frac1{2\pi\alpha'}\partial_\tau X^7,\nonumber\\
\Pi_8&=&\frac1{2\pi\alpha'}(\partial_\tau X^8 - 2\tilde\mu X^7).
\end{eqnarray}

The mode expansions are straightforwardly obtained  \cite{Mic}.
Equations of motion are
\begin{eqnarray}
&&\stackrel{..~}{X^I} - {X^I}''+\tilde\mu^2X^I =0, \nonumber \\
&&\stackrel{..~}{X^7} - {X^7}''+2\tilde\mu\stackrel{.~}{X^8} =0, \\
&&\stackrel{..~}{X^8} - {X^8}''-2\tilde\mu\stackrel{.~}{X^7} =0.
\nonumber
\end{eqnarray}
With the boundary condition
\begin{eqnarray}
X^I(\tau,\sigma=2\pi)=X^I(\tau,\sigma=0)~~~(I=1,\ldots,6),
\end{eqnarray}
$X^I$'s are expanded as
\begin{eqnarray}
X^I(\tau,\sigma)
&=&x_0^I\cos \tilde\mu\tau + \frac{\alpha'}{\tilde\mu}p^I \sin
\tilde\mu\tau
+i\sqrt{\frac{\alpha'}2}\sum_{n\neq 0}
\frac1n\left(
a^I_n f_n^+
+\tilde a^I_n f_n^-
\right),\label{X^I}
\end{eqnarray}
where
\begin{eqnarray}
f_n^{\pm}(\tau,\sigma)&\equiv& e^{-in(\rho_n\tau\pm\sigma)},\label{fn}\\
\rho_n & \equiv& \sqrt{1+\frac{\tilde{\mu}^2}{n^2}}.\nonumber
\end{eqnarray}
The commutation relations are
\begin{eqnarray}
[x_0^I,~p^J]&=&i\delta^{IJ},\nonumber\\
{[}a_n^I,~a_m^J{]}&=&{[}\tilde a_n^I,~\tilde a_m^J{]}~=~
\frac n{\rho_n }\delta^{IJ}\delta_{n,-m}.
\end{eqnarray}
\indent
$X^7$ and $X^8$ have similar expansions to $X^I$ since
$Y\equiv e^{\tilde\mu i \tau}(X^8+ i X^7)$ satisfies the same equation
of motion
as $X^I$. Their boundary conditions are
\begin{eqnarray}
&&X^7(\tau,\sigma=2\pi)=X^7(\tau,\sigma=0),~~~
X^8(\tau,\sigma=2\pi)=X^8(\tau,\sigma=0)+2\pi R_{IIB}w^8
\end{eqnarray}
where $w^8\in\mbox{\boldmath $Z$}$ is the winding number.
Writing $X\equiv X^8+iX^7$, $\bar X\equiv X^8-iX^7$, they are
\begin{eqnarray}
X(\tau,\sigma)&=&
x_0^8+i x_0^7
+ R_{IIB}w^8\sigma
+a_0\sqrt{\frac{\alpha'}{\tilde\mu}} e^{-2i\tilde\mu \tau}
+e^{-i\tilde\mu \tau}i\sqrt{\frac{\alpha'}2}
\sum_{n\neq 0}
\frac1n\left(
a_n f_n^+
+\tilde a_n f_n^-
\right),\nonumber\\
\bar X(\tau,\sigma)&=&
x_0^8-i x_0^7
+ R_{IIB}w^8\sigma
+\bar a_0\sqrt{\frac{\alpha'}{\tilde\mu}} e^{2i\tilde\mu \tau}
+e^{i\tilde\mu \tau}i\sqrt{\frac{\alpha'}2}
\sum_{n\neq 0}
\frac1n\left(
\bar a_n f_n^+
+\bar{\tilde a}_n
f_n^-
\right)
\label{XXbar}
\end{eqnarray}
with
\begin{eqnarray}
{[}x_0^7,~x_0^8{]}&=&\frac{\alpha' i}{2\tilde\mu} \label{x078}
\end{eqnarray}
and
\begin{eqnarray}
{[}a_0,~\bar a_0{]}&=&1, \nonumber \\
{[}a_n,~\bar a_m{]}&=&{[}\tilde a_n,~\bar{\tilde a}_m{]}~=~
\frac {2n}{\rho_n }\delta_{n,-m} ~~~(n,m\neq
0),
\label{a_ncommutators}
\end{eqnarray}
where the zero-mode part is different from \cite{Mic} but is
equivalent.
Note that, unlike the ordinary toroidal case, the compact boson $X^8$
does not
have quantized
momenta owing to ``Coriolis's force" in the 7-8 plane.\footnote{
This is a realization of a noncommutative cylinder without (tangent)
constant $B$ field (see (\ref{x078})).
}
    Therefore the  ``Narain lattice"  is
a one-dimensional lattice here.

\noindent{\it IIA spectrum}\\
We next consider the IIA background obtained by the T-duality
transformation
along $\partial/\partial X^8$ \cite{AGGP},
which can be read off from the reduction to nine dimensions. The metric
and non-vanishing
components of other fields are
\begin{eqnarray}
&&ds_{IIA}^2 ~=~ 2dX^+dX^- - \mu^2 ((X^I )^2+4(X^7)^2) (dX^+)^2 + dX^i
dX^i,\nonumber\\
&&F_{4~+567}~=~4\mu,~~~H_{+79}~=~-2\mu
\end{eqnarray}
with $i=1,\ldots,7,9$ and $I=1,\ldots,6$,
where $X^9$ denotes the coordinate of the dual circle. All other
components vanish.
The bosonic light-cone string action reads
     \begin{eqnarray}
S_{IIA}&=&\int d\tau L_{IIA}, \nonumber\\
L_{IIA}&=&
\int_0^{2\pi} d\sigma
\left[
p^+\partial_\tau X^-
+\frac1{4\pi\alpha'}
\left(
(\partial_\tau X^i)^2-(\partial_\sigma X^i)^2
    \right.\right.\nonumber\\ &&~~~~~~~~~~~\left.\left.
-\tilde\mu^2 ((X^I)^2+4(X^7)^2)
-4\tilde\mu X^7\partial_\sigma X^9
\rule{0mm}{5mm}\right)
\right].
\end{eqnarray}
The canonical momenta $\Pi_I$ and $\Pi_7$ are the same as IIB, and
\begin{eqnarray}
\Pi_9&=&\frac1{2\pi\alpha'}\partial_\tau X^9.
\end{eqnarray}
The Hamiltonian is
\begin{eqnarray}
H_{IIA}&=&\frac1{4\pi\alpha'}\int_0^{2\pi}d\sigma\left[
(2\pi\alpha')^2
\Pi_i^2
+(\partial_\sigma X^i)^2
+\tilde\mu^2 ((X^I)^2+4(X^7)^2) +4\tilde\mu X^7\partial_\sigma X^9
\rule{0mm}{5mm}\right].
\end{eqnarray}

Since the $X^I$ equations of motion are identical to those before
taking T-dual,
their mode expansions remain the same. The $X^7$ and $X^9$ equations
are
\begin{eqnarray}
&&\stackrel{..~}{X^7} - {X^7}''+4\tilde\mu^2X^7 +2\tilde\mu {X^9}'=0,
   \nonumber \\
&&\stackrel{..~}{X^9} - {X^9}'' ~~~~~~~~~~~~ -2\tilde\mu {X^7}'=0
\end{eqnarray}
and their boundary conditions are
\begin{eqnarray}
&&X^7(\tau,\sigma=2\pi)=X^7(\tau,\sigma=0),~~~
X^9(\tau,\sigma=2\pi)=X^9(\tau,\sigma=0)+2\pi R_{IIA}w^9.
\end{eqnarray}
$w^9\in\mbox{\boldmath $Z$}$ is the winding number.
If $(X^7, X^9)$ is a solution with $w^9=0$, so is
$(X^7-\frac{R_{IIA}}{2\tilde\mu}w^9, X^9+R_{IIA}w^9\sigma)$
with winding number $w^9$. All the winding solutions are obtained in
this way.
Therefore it is enough to consider solutions with $w^9 =0$.

For nonzero modes, we make an ansatz :
\begin{eqnarray}
X^7(\tau,\sigma)&=&A^7 e^{i(\omega_n\tau + n\sigma)},\nonumber\\
X^9(\tau,\sigma)&=&A^9 e^{i(\omega_n\tau + n\sigma)}.
\end{eqnarray}
$A^7$, $A^9$ and $\omega_n$ are constants $(n\neq 0)$.
Plugging them into the equations of motion, one finds that the
solutions satisfy
either
\begin{eqnarray}
&&\omega_n~=~\tilde\mu\pm n\rho_n ,~~~~~
\frac{A^9}{A^7}~=~\frac{-\tilde\mu\pm
n\rho_n }{in}
\end{eqnarray}
or
\begin{eqnarray}
&&\omega_n~=~-\tilde\mu\pm n\rho_n ,~~~~~
\frac{A^9}{A^7}~=~\frac{-\tilde\mu\mp
n\rho_n }{in}.
\end{eqnarray}
On the other hand, the modes independent of $\sigma$ satisfy
\begin{eqnarray}
&&\stackrel{..~}{X^7} +4\tilde\mu^2X^7 =0, ~~~~~\stackrel{..~}{X^9}  =0.
\end{eqnarray}
Taking into account the winding modes together, we obtain the expansions
\begin{eqnarray}
    X^7_{IIA}&=&-\frac{R_{IIA}}{2\tilde\mu}w^9
+\frac1{2i} \sqrt{\frac{\alpha'}{\tilde\mu}}
(a_0 e^{-2\tilde\mu i \tau} - \bar a_0 e^{2\tilde\mu i \tau})
     \label{X7IIA} \\
    && +\frac 12 \sqrt{\frac{\alpha'}2}e^{-i\tilde\mu \tau}
\sum_{n\neq 0}\frac1n\left(
a_n f_n^+
+\tilde a_n f_n^-
\right)
-\frac 12 \sqrt{\frac{\alpha'}2}e^{i\tilde\mu \tau}
\sum_{n\neq 0}\frac1n\left(
\bar a_n f_n^+
+\bar{\tilde a}_n  f_n^-
\right), \nonumber
\\
X^9 &=& R_{IIA}w^9\sigma + x^9_0 + \alpha' p^9 \tau \\
&& +\frac 12 \sqrt{\frac{\alpha'}2}e^{-i\tilde\mu \tau}
\sum_{n\neq 0}\frac{\tilde\mu\!-\!n\rho_n }{in^2}
\left(a_n f_n^+-\tilde a_n f_n^-\right)
-\frac 12 \sqrt{\frac{\alpha'}2}e^{i\tilde\mu \tau}
\sum_{n\neq 0}\frac{\tilde\mu\!+\!n\rho_n }{in^2}
\left( \bar a_n f_n^+ -\bar{\tilde a}_n f_n^-\right) \nonumber
\end{eqnarray}
with ${[} x^9_0~,~p^9{]}=i$ and the same relations as
(\ref{a_ncommutators}).

\noindent{\it Comparison of the spectra}\\
We have put the subscript ``{\it IIA}" on $X^7$ in order to distinguish
from the
IIB expansion, though we have used the same oscillators; this is
justified since
$X^7_{IIA}$ (\ref{X7IIA}) coincides with $X^7=(2i)^{-1}(X-\bar X) $
(\ref{XXbar}) up to constant modes, of which the Hamiltonians are
independent.\footnote{Due to the quantization of
$x_0^7$ discussed below, the constant modes also agree if
the parameters are appropriately identified.}
Therefore, just like the ordinary toroidal
compactification, no
distinction is necessary between $X^7$ and $X^7_{IIA}$.

On the other hand, unlike $X^8$, the ``dual" boson $X^9$ has not only
a winding mode but also a momentum mode, since the equation of motion
for the $X^9$ zero modes is free and the same as the flat case. One can
argue the quantization of this momentum by noting that $p^9$ is a
constant
part of the canonical momentum $2\pi \Pi_9$ (or by requiring the
solution
to the Klein-Gordon equation $e^{ip^9X^9}$ to be single-valued),
to get
\begin{eqnarray}
p^9&=&\frac{n^9}{R_{IIA}}~~~(n^9\in\mbox{\boldmath $Z$}).
\end{eqnarray}
Therefore, the Narain lattice on the ``dual" background appears to
be spanned by $n^9$ and $w^9$, and hence does not look like dual to the
one-dimensional lattice for $X^8$. However, an explicit computation
shows that the $w^9$ dependences cancel in the Hamiltonian, and the
lattice
becomes again one-dimensional.
More precisely, one may confirm that the
following relations hold:
\begin{eqnarray}
2\pi\alpha' \Pi_8~~(IIB)&=&\partial_\sigma X^9~~(IIA),\nonumber\\
2\pi\alpha' \Pi_9~~(IIA)&=&\partial_\sigma X^8~~(IIB), \label{Pi=dX}
\end{eqnarray}
provided that\footnote{Here we have used the scheme in which the metric
is kept
fixed but the radius $R$ describes the change of the physical
circumference.
In the next section we fix the radii but get the metric transformed.
}
\begin{eqnarray}
w^8=n^9,~~~R_{IIB}=\frac{\alpha'}{R_{IIA}},
\label{w8=n9}
\end{eqnarray}
and then
\begin{eqnarray}
H_{IIB}=H_{IIA}.
\end{eqnarray}
We note that, on the IIA side, there is an infinite degeneracy of the 
spectrum in the Hilbert space  associated with the redundancy
of $w^9$. The corresponding degeneracy on the IIB side comes
from the noncommutative coordinate $x_0^7$, which is the constant 
piece of $2\pi \Pi_8$ and, hence, quantized as
$ - \alpha' n^8/(2 \tilde{\mu} R_{IIB}) $.\footnote{
We thank Y.~Imamura and T.~Yoneya for comments on this point.
See also \cite{BBHIO}.}

To compare the physical spectra, we have to consider 
the constraints, details of which for plane-wave backgrounds are found
in \cite{Met}. The physical spectrum then needs to satisfy 
the ``level-matching'' condition:
\begin{eqnarray}
  P \equiv \int_0^{2\pi} d\sigma \Pi_i \partial_\sigma X^i = 0.
\end{eqnarray}
The oscillator part of $P$ is computed straightforwardly,
while the quantization conditions for the zero modes give 
$P_{IIB}^{\rm zeromodes} = n^8 w^8$,
$P_{IIA}^{\rm zeromodes} = n^9 w^9$. 
If $w^8 = n^9 $ and $n^8=w^9$, 
\begin{eqnarray}
P_{IIB}=P_{IIA}
\end{eqnarray}
holds, and the physical states also correspond one to one.
This establishes that this T-dual pair is ``really dual".
Note that,  when $w^8 \neq 0 \, (n^9 \neq 0) $, 
the physical-state condition resolves 
the infinite degeneracy in the Hilbert space and 
selects a unique $ n^8 (w^9)$ for the physical states.   

It would be worth noting that the relations (\ref{Pi=dX}) directly
follow from the usual
duality transformation \cite{B} (introducing $X^9$ as a Lagrange
multiplier and
integrating out $X^8$) in  the string action (\ref{IIBstring}), and
the classical coincidence of the Hamiltonians follows.
Our direct check shows that it is also consistent quantum mechanically,
mode by mode, in particular that the Narain lattices agree despite the
apparent asymmetry of zero modes.

\noindent{\it $T^d$ compactifications $(d=1,\ldots,4)$}\\
The coordinate system (\ref{coordinates}) can be trivially generalized
so that
up to four commuting Killing vectors can be manifestly seen. For
instance, the Killing vectors
$k_{S^+_{21}}$,$k_{S^+_{43}}$,$k_{S^+_{65}}$ and $k_{S^+_{87}}$ are
manifest in the
coordinate system
\begin{eqnarray}
x^+&=& X^+,\nonumber\\
x^-&=& X^- -\mu (X^1X^2+\cdots+X^7X^8),\nonumber\\
\left[\begin{array}{c}x^{2j-1}\\x^{2j}\end{array}\right]&=&
\left[\begin{array}{rc}\cos \mu X^+&\sin\mu X^+ \\
-\sin\mu X^+&\cos\mu X^+\end{array}\right]
\left[\begin{array}{c}X^{2j-
1}\\X^{2j}\end{array}\right]~~~~~~(j=1,\ldots,4).
\label{coordinates2}
\end{eqnarray}
The metric then reads
\begin{eqnarray}
ds^2 &=& 2dX^+dX^-  -4\mu (X^1 dX^2 +\cdots +X^7 dX^8) dX^+ + dX^i dX^i
~~~~(i=1,\ldots,8).
\end{eqnarray}
With this choice of a set of Killing vectors, $F_{5}$ is again
unchanged.

In this case, the transverse boson mode expansions are nothing but four
copies
of ($X^7$, $X^8$) (\ref{XXbar}) in the $S^1$ case, and hence contain a
four-dimensional
winding-number lattice but no momentum lattice. Then a simultaneous
T-duality flip along
$X^2$, $X^4$, $X^6$ and $X^8$ directions converts it to a
four-dimensional momentum
lattice.

\section{$O(2,2; \mbox{\boldmath $Z$})$ symmetry}
In $T^d$ compactifications for ($d\geq 2$), the simultaneous T-duality
flip
we have considered so far is a special element of
$O(d,d;\mbox{\boldmath $Z$})$
transformation, realized by conjugating the $O(d,d)$ metric $L$ to the
scalar
matrix. We will now examine the full $O(2,2;\mbox{\boldmath $Z$})$
duality
in the plane-wave background compactified on $T^2$.

The metric that we use for a $T^2$ compactification is
\begin{eqnarray}
ds^2 &=& 2dX^+dX^- -\mu^2X^I X^I(dX^+)^2
-4\mu (X^5 dX^6 +X^7 dX^8) dX^+ + dX^i dX^i
\end{eqnarray}
($i=1,\ldots,8$; $I=1,\ldots,4$). The dilaton and $B$ fields are zero.
We have taken $k_{S^+_{65}}$ and $k_{S^+_{87}}$
as the manifest Killing vectors. The change of coordinates is an obvious
generalization of (\ref{coordinates}) similar to (\ref{coordinates2})
\cite{Mic}.
$F_5$ is the same as before but anyway irrelevant for the bosonic
spectrum.

We compactify the manifest coordinates $X^6$, $X^8$ as
\begin{eqnarray}
X^6\sim X^6+2\pi R^6, ~~~X^8\sim X^8+2\pi R^8,
\end{eqnarray}
then the $O(2,2;\mbox{\boldmath $Z$})$ transformation $g$ acting on the
scalar matrix $\cal M$ as $g{\cal M}g^T$ is given by
\begin{eqnarray}
g=Q^{-1}g_{\mbox{\scriptsize\boldmath $Z$ }}Q,~~~
Q=\left[
\begin{array}{cccc}
R^6/\alpha'&&&\\
&R^8/\alpha'&&\\
&&1/R^6&\\
&&&1/R^8
\end{array}
\right],
\end{eqnarray}
where $g_{\mbox{\scriptsize\boldmath $Z$ }}$ is an
$O(2,2;\mbox{\boldmath $R$})$ matrix with integer entries.
Any $g_{\mbox{\scriptsize\boldmath $Z$ }}$ can be written
\cite{GPR} as
$g_{\mbox{\scriptsize\boldmath $Z$ }}=w^s g_1  g_2$,
where
\begin{eqnarray}
w=\left[
\begin{array}{rrrr}
1&&&\\
&&&1\\
&&1&\\
&1&&
\end{array}
\right],~~~ s=0 ~\mbox{or}~1,
\end{eqnarray}
and $g_1\in G_1$, $g_2\in G_2$ with
$G_1,G_2=SL(2,\mbox{\boldmath $Z$})$ respectively represented by the
matrices of the form
\begin{eqnarray}
g_1=\left[
\begin{array}{rrrr}
a&b&& \\
c&d&&\\
&&d&-c\\
&&-b&a
\end{array}
\right],~~~
g_2=\left[
\begin{array}{rrrr}
a&&&b \\
&a&-b&\\
&-c&d&\\
c&&&d
\end{array}
\right].
\end{eqnarray}
$a,b,c,d$ are integers satisfying $ad-bc=1$.
We have  taken the
$O(d,d)$ metric $L$ as
\begin{eqnarray}
L=\left[
\begin{array}{rrrr}
&&1& \\
&&&1\\
1&&&\\
&1&&
\end{array}
\right],
\end{eqnarray}
so that $gLg^T=L$.

How the $O(2,2; \mbox{\boldmath $Z$})$ acts on a supergravity
background may
be read off from the dimensionally reduced NS-NS sigma model
\cite{MS}.
It is easy to see that $G_1$, one $SL(2,\mbox{\boldmath $Z$})$ factor of
$O(2,2; \mbox{\boldmath $Z$}$), causes no change in the spectrum, since
it can be absorbed into the target space modular transformation of
compactified
($X^6$,$X^8$). $w$ is a T-duality flip along a single circle, and the
invariance
of the Hamiltonian was confirmed in the previous section. Therefore, we
have
only to prove the invariance under the action of $G_2$ and $w G_1 
~(=G_2 w)$.
First we check $G_2$.

Writing
\begin{eqnarray}
\tilde{b} = \frac{R^6R^8}{\alpha'} b,~~~
\tilde{c} = \frac{\alpha'}{R^6R^8} c,
\end{eqnarray}
we obtain the new metric in the form
\begin{eqnarray}
ds_{\mbox{\scriptsize new}}^2&=&
2dX^+dX^- -\mu^2\left(X^I X^I
+\frac{4\tilde{c}^2}{\tilde{c}^2+d^2}X^{\hat m}X^{\hat m}\right)(dX^+)^2
\\
&&-\frac{4d\mu}{\tilde{c}^2+d^2} (X^5 dX^6 +X^7 dX^8) dX^+ + dX^I dX^I
+dX^{\hat m}dX^{\hat m}
+\frac1{\tilde{c}^2+d^2}dX^m dX^m,\nonumber
\end{eqnarray}
and nonvanishing components of the $B$ field
\begin{eqnarray}
B^{\mbox{\scriptsize
new}}_{+6}&=&\frac{2\tilde{c}\mu}{\tilde{c}^2+d^2}X^7,
\quad
B^{\mbox{\scriptsize new}}_{+8}
   \ = \ -\frac{2\tilde{c}\mu}{\tilde{c}^2+d^2}X^5, \quad
B^{\mbox{\scriptsize new}}_{68}
   \ = \ \frac{a\tilde{c}+\tilde{b}d}{\tilde{c}^2+d^2}.
\end{eqnarray}
The bosonic string action  reads
\begin{eqnarray}
S^{\mbox{\scriptsize new}}&=&\int d\tau L^{\mbox{\scriptsize new}},
\\
L^{\mbox{\scriptsize new}}&=&
\int_0^{2\pi} d\sigma
\left[
p^+\partial_\tau X^-
+\frac1{4\pi\alpha'}
\left(
(\partial_\tau X^I)^2-(\partial_\sigma X^I)^2
     -\tilde\mu^2 \left((X^I)^2
   +\frac{4\tilde{c}^2}{\tilde{c}^2+d^2}(X^{\hat m})\right)
\right.\right.
\nonumber\\&&
~~~~~~~~~~~~~~~~~~~~~+(\partial_\tau X^{\hat m})^2-(\partial_\sigma
X^{\hat m})^2
+\frac1{\tilde{c}^2+d^2}(
(\partial_\tau X^m)^2-(\partial_\sigma X^m)^2)
\nonumber\\&&
~~~~~~~~~~~~~~~~~~~~~
-\frac{2(a\tilde{c}+\tilde{b}d)}{\tilde{c}^2+d^2}
(\partial_\tau X^6\partial_\sigma X^8
-\partial_\tau X^8\partial_\sigma X^6)
\nonumber\\&&\left.\left.
~~~~~~~~~~~~~~~~~~~~~
-\frac{4\tilde\mu }{\tilde{c}^2+d^2}
(d(X^5\partial_\tau X^6+X^7\partial_\tau X^8)
+\tilde{c}(X^7\partial_\sigma X^6-X^5\partial_\sigma X^8))
\rule{0mm}{5mm}\right)
\right],\nonumber
\end{eqnarray}
where $\hat m=5,7$ and $m=6,8$. The $X^I$ modes are
inert under the
$SL(2,\mbox{\boldmath $Z$})$,
while the $X^{\hat m}$ and $X^m$ equations of motion are
\begin{eqnarray}
&&\stackrel{..~}{X^5} - {X^5}''+\frac{2\tilde\mu}{\tilde{c}^2+d^2}
(d\stackrel{.~}{X^6}-\tilde{c}{X^8}'+2\tilde\mu \tilde{c}^2 X^5) =0,
\nonumber \\
&&\stackrel{..~}{X^7} - {X^7}''+\frac{2\tilde\mu}{\tilde{c}^2+d^2}
(d\stackrel{.~}{X^8}+\tilde{c}{X^6}'+2\tilde\mu \tilde{c}^2 X^7) =0,
\nonumber \\
&&\stackrel{..~}{X^6} -
{X^6}''-2\tilde\mu(d\stackrel{.~}{X^5}+\tilde{c}{X^7}'
)=0, \nonumber \\
&&\stackrel{..~}{X^8} -
{X^8}''-2\tilde\mu(d\stackrel{.~}{X^7}-\tilde{c}{X^5}')
=0.
\end{eqnarray}
It can be verified that, for the solutions of the form
$e^{i(\omega_n\tau+n\sigma)}$, $\omega_n$ does not
depend on $c $ and $d$, but is always given by
\begin{eqnarray}
\omega_n=\pm\tilde\mu\pm\sqrt{\tilde\mu^2+n^2}.
\end{eqnarray}
Therefore, the set of nonzero modes is $SL(2,\mbox{\boldmath $Z$})$
independent. On the other hand, the zero modes are found to be
\begin{eqnarray}
X^5_{\mbox{\scriptsize zeromodes}}&=&\frac1{2\tilde\mu
\tilde{c}^2}(-d\alpha'p^6+\tilde{c}R^8w^8),\nonumber\\
X^6_{\mbox{\scriptsize
zeromodes}}&=&x_0^6+\alpha'p^6\tau+R^6w^6\sigma,\nonumber\\
X^7_{\mbox{\scriptsize zeromodes}}&=&\frac1{2\tilde\mu
\tilde{c}^2}(-d\alpha'p^8-\tilde{c}R^6w^6),\nonumber\\
X^8_{\mbox{\scriptsize zeromodes}}&=&x_0^8+\alpha'p^8\tau+R^8w^8\sigma
\end{eqnarray}
if $c\neq 0$.
They depend both on the momenta $p^m$ and on the winding numbers
$w^m\in\mbox{\boldmath $Z$}$.
Using $f_n^{\pm}(\tau,\sigma)$ (\ref{fn}),
we obtain the mode expansions
\def\mut{\tilde{\mu}}
\def\mhat{\hat{m}}
\begin{eqnarray}
X^{\mhat}(\tau,\sigma)&=&x_0^{\mhat}+y^{\mhat}
+\frac1{2i}\sqrt{\frac{\alpha'}{\mut}}
(a_0^{\mhat}e^{-2i\mut\tau}-\bar a_0^{\mhat}e^{2i\mut\tau})\\
&&+\frac12\sqrt{\frac{\alpha'}2}e^{-i\mut\tau}
\sum_{n\neq 0}\frac1n\left(a_n^{\mhat} f_n^+
+\tilde a_n^{\mhat} f_n^-\right)
-\frac12\sqrt{\frac{\alpha'}2}e^{i\mut\tau}
\sum_{n\neq 0}\frac1n\left(\bar{a}_n^{\mhat} f_n^+
    +\bar{\tilde a}_n^{\mhat} f_n^-\right),
\nonumber\\
y^{\mhat}&=&\left\{\begin{array}{ll}
+\frac{R^8w^8}{2\mut \tilde{c}}&(\mhat=5), \\
-\frac{R^6w^6}{2\mut \tilde{c}}&(\mhat=7),
\end{array}\right.
\nonumber\\
X^6(\tau,\sigma)&=&x_0^6+\alpha'p^6\tau + R^6w^6\sigma
+\frac12\sqrt{\frac{\alpha'}{\mut}}
d(a_0^5 e^{-2i\mut\tau}+\bar a_0^5 e^{2i\mut\tau})\\
&&+\frac i2\sqrt{\frac{\alpha'}2}e^{-i\mut\tau}
\sum_{n\neq 0}\frac1n\left((da_n^5-\frac{\tilde{c}
(\mut\!-\!n\rho_n)}n a_n^7) f_n^+
+(d\tilde{a}_n^5+\frac{\tilde{c}(\mut\!-\!n\rho_n)}n \tilde{a}_n^7)
f_n^-\right)
\nonumber\\
&&-\frac i2\sqrt{\frac{\alpha'}2}e^{i\mut\tau}
\sum_{n\neq 0}\frac1n\left((-d{\bar a}_n^5
-\frac{\tilde{c}(\mut\!+\!n\rho_n)}n \bar{a}_n^7) f_n^+
+(-d{\bar{\tilde a}}_n^5+
\frac{\tilde{c}(\mut\!+\!n\rho_n)}n \bar{\tilde a}_n^7) f_n^-\right),
\nonumber\\
X^8(\tau,\sigma)&=&x_0^8+\alpha'p^8\tau + R^8w^8\sigma
+\frac12\sqrt{\frac{\alpha'}{\mut}}
d(a_0^7 e^{-2i\mut\tau}+\bar a_0^7 e^{2i\mut\tau})\\
&&  +\frac i2\sqrt{\frac{\alpha'}2}e^{-i\mut\tau}
\sum_{n\neq 0}\frac1n\left((da_n^7+\frac{\tilde{c}(\mut\!-\!n\rho_n)}n
a_n^5)
f_n^+
+ (d\tilde{a}_n^7-\frac{\tilde{c}(\mut\!-\!n\rho_n)}n \tilde{a}_n^5)
f_n^-\right)
\nonumber\\
&&  -\frac i2\sqrt{\frac{\alpha'}2}e^{i\mut\tau}
   \sum_{n\neq 0}\frac1n\left((-d{\bar a}_n^7
+\frac{\tilde{c}(\mut\!+\!n\rho_n)}n \bar{a}_n^5) f_n^+
+(-d{\bar{\tilde a}}_n^7
-\frac{\tilde{c}(\mut\!+\!n\rho_n)}n \bar{\tilde a}_n^5) f_n^-\right)
\nonumber
\end{eqnarray}
with the relations $d\alpha' p^6=-2\mut \tilde{c}^2x_0^5$,
$d\alpha' p^8=-2\mut \tilde{c}^2x_0^7$.
The expansions of $X^I$'s $(I=1,...,4)$ are the same as (\ref{X^I}).
The relevant commutation relations are
\begin{eqnarray}
{[}x_0^m,~{p^{m'}}{]}&=&i\tilde{c}^2 \delta^{mm'}~~~(m,m'=6,8)
~~~(\mbox{or}~~{[}x_0^5,~x_0^6{]}~=~{[}x_0^7,~x_0^8{]}~
=~i\frac{d\alpha'}{2\mut}), \nonumber\\
{[}a_0^{\mhat},~\bar a_0^{\mhat'}{]}&=&
\delta^{\mhat \mhat'}, \nonumber \\
{[}a_n^{\mhat},~\bar a_l^{\mhat'}{]}&=&
{[}\tilde a_n^{\mhat},~\bar{\tilde a}_l^{\mhat'}{]}~=~
\delta^{\mhat \mhat'}\delta_{n,-l}\frac{2n}{\rho_n}. \label{comzeros}
\end{eqnarray}
Plugging these mode expansions into the Hamiltonian and 
world-sheet momentum, we have confirmed
after some calculation that the contributions from the nonzero modes
are indeed
invariant under $SL(2,\mbox{\boldmath $Z$})$.
The contributions from the zero modes to the Hamiltonian
$H_{\mbox{\scriptsize zeromodes}}$
turn out to be
\begin{eqnarray}
H_{\mbox{\scriptsize
zeromodes}}=\frac{\alpha'}{2\tilde{c}^2}\left((p^6)^2+(p^8)^2\right).
\end{eqnarray}
At first sight it appears to depend on $c$. However, the constant
pieces of
$2\pi \Pi_m$ $(m=6,8)$ are
\begin{eqnarray}
\frac{p^6}{\tilde{c}^2}-\frac{a}{\tilde{c}} \frac{R^8}{\alpha'} w^8,~~~
\frac{p^8}{\tilde{c}^2}+\frac{a}{\tilde{c}} \frac{R^6}{\alpha'} w^6,
\end{eqnarray}
respectively, which imply the quantization conditions
\begin{eqnarray}
\frac{p^6}{\tilde{c}}&=& \frac{R^8}{\alpha'} w'^8, \quad
\frac{p^8}{\tilde{c}} \ = \ -\frac{R^6}{\alpha'} w'^6,
\end{eqnarray}
where
\begin{eqnarray}
   \left[\begin{array}{c}w'^8\\n'^6\end{array}\right]&=&
\left[\begin{array}{rc} a & c \\
   b &d\end{array}\right]
\left[\begin{array}{c}w^8\\n^6\end{array}\right], \qquad
   \left[\begin{array}{c}w'^6\\n'^8\end{array}\right] \ = \
\left[\begin{array}{rc} a & -c \\
   -b &d\end{array}\right]
\left[\begin{array}{c}w^6\\n^8\end{array}\right]
   \label{quantp}
\end{eqnarray}
with $n^6, n^8 \in \mbox{\boldmath $Z$} $.
Since ${w'}^m$,${n'}^m$ ($m=6,8$)  run over
$\mbox{\boldmath $Z$}$,
$H_{\mbox{\scriptsize zeromodes}}$ is invariant under the
$SL(2,\mbox{\boldmath $Z$})$. Similarly, we find that 
\begin{eqnarray}
P_{\mbox{\scriptsize zeromodes}}= n'^6 w'^6 + n'^8 w'^8, \label{P}
\end{eqnarray}
and $P_{\mbox{\scriptsize zeromodes}}$ is also invariant.

Finally, if $c=0$, the zero modes are simply given by
\begin{eqnarray}
   && X^{5}_{\mbox{\scriptsize zeromodes}} = x_0^{5}, \qquad
      X^{6}_{\mbox{\scriptsize zeromodes}}
       = x_0^{6} + R^{6}w^{6} \sigma, \nonumber \\
    && X^{7}_{\mbox{\scriptsize zeromodes}} = x_0^{7}, \qquad
      X^{8}_{\mbox{\scriptsize zeromodes}}
       = x_0^{8} + R^{8}w^{8} \sigma,
\end{eqnarray}
where the  commutation relations of $(x_0^{\hat{m}},x_0^m)$
are the same as in (\ref{comzeros}).
The spectrum then amounts to two sets of the IIB
spectrum found in the previous section, with constant shifts of the
canonical momenta. The invariance of the spectrum
is easily confirmed. Thus we have shown the $G_2$ invariance.

One can similarly check the $w G_1 ~(=G_2 w)$ invariance. In this case
much labor is saved because the system after the transformation is
reduced to that of the IIB bosons $X^7,X^8$ and the IIA bosons
$X^7_{IIA},X^9$ in the $S^1$ compactification,
with a constant compact  $B$ field. Then the check of the invariance 
for the
nonzero-mode sector is trivial, and for the zero-mode sector is also 
straightforward.
This completes our proof of the $O(2,2; \mbox{\boldmath $Z$})$
T-duality in the $T^2$ compactified IIB maximally supersymmetric plane
wave.

\section{Conclusions and discussion}
We have shown, by direct computations of bosonic string spectra,
the $O(d,d; \mbox{\boldmath $Z$})$ $(d=1,2)$ T-duality
in the maximally supersymmetric IIB plane-wave background
compactified on $S^1$ and $T^2$.
Only half of the ordinary set of zero modes appear in the Hamiltonian.
The ``half" Narain lattice is stabilized by the T-duality group.

A natural question may be raised here about modular invariance.
Let us focus on the case of the action of $G_2$. Other cases are
similar.
First, the partition function is given by
\begin{eqnarray}
    Z & \sim &
      \, {\rm Tr}\, e^{-2\pi \tau_2 (\alpha' p^+p^- + H )
+ 2 \pi i \tau_1 P} , \label{partition}
\end{eqnarray}
where $\tau_1+i\tau_2$ is the modular parameter.
Since the zero-mode piece of $P$ 
is given by (\ref{P}) for both $c=0$ and $c\neq 0$ and 
the Hamiltonian is independent of $(n'^6,n'^8)$, the summation
over these integers in the trace
gives $\delta_{w'^6,0} \delta_{w'^8,0}$.
(Precisely, this holds for irrational $\tau_1$. The case of rational
$\tau_1$ may be safely handled, since its measure is zero.)
This in turn makes the summation over $(w'^6,w'^8)$ trivial.
Thus, modular invariance does not require pairs of integers
in the Hamiltonian, which is different from the flat case.
In this way,
mechanism of T-duality works in an interesting manner, not to make
the ``half" Narain lattice conflict with modular invariance.
(For a related discussion in a somewhat different context,
see \cite{Sugawara}.)
For the uncompactified plane-wave background (\ref{noncompactpp}),
the modular invariance of the one-loop vacuum amplitude has been
discussed in \cite{Takayanagi-Bergman-GG}.
It would be an interesting open problem to show
the modular invariance also in the compactified cases.

The background we have considered has more commuting isometries. For
example,
\begin{eqnarray}
\{
k_{S^+_{21}}, k_{S^+_{43}}, k_{S^+_{65}}, k_{S^+_{87}},
k_{S^+_{23}}, k_{S^+_{45}}, k_{S^+_{67}}, k_{S^+_{81}}
\}
\end{eqnarray}
are a (non-maximal) set of mutually commuting Killing vectors
with unit norm. The first and
last four's are mutually orthogonal, but some two of the whole set
are not.
Therefore, while it is natural to expect that the T-duality group
extends at least
up to $O(4,4;\mbox{\boldmath $Z$})$, it  would be interesting to see
whether
and how it goes beyond that.

Finally, one of the underlying motivations for the present work is to
investigate
whether  $E_{10}$ \cite{Jul} is the duality of the maximally
supersymmetric
plane wave. Being a one-dimensional system with a null Killing vector
    (in the Rosen coordinates), it is a solution to the null reduction to
one dimension
\cite{Nic-Miz}. Since both are, in a sense, something ultimate, it is
tempting
to speculate that the maximally supersymmetric plane wave might realize
$E_{10}$ as its U-duality.

\vspace{4ex}
\section*{Acknowledgments}
We would like to thank Y.~Imamura, N.~Ishibashi, Y.~Sugawara,
S.~Terashima and T.~Yoneya for discussions.
The work of S.M. is supported in part by Grant-in-Aid
for Scientific Research (C)(2) \#14540286 from
The Ministry of Education, Culture, Sports, Science
and Technology, whereas the work of Y.S. is supported in part
by University of Tsukuba Research Projects.

Y.S. would like to dedicate this paper to the memory of Youngjai Kiem.

\newpage
%

%
\end{document}